\documentclass[a4paper,fleqn]{cas-sc}

\usepackage[authoryear,longnamesfirst]{natbib}

\usepackage[caption=false]{subfig}
\usepackage{amsmath}

\def\tsc#1{\csdef{#1}{\textsc{\lowercase{#1}}\xspace}}
\tsc{WGM}
\tsc{QE}
\tsc{EP}
\tsc{PMS}
\tsc{BEC}
\tsc{DE}
\begin{document}
\let\WriteBookmarks\relax
\def\floatpagepagefraction{1}
\def\textpagefraction{.001}

%\shorttitle{Enhancing M\&A Predictive Modeling from patent data}

\shorttitle{ML-based M\&A prediction from patent data}

\shortauthors{G Albora et~al.}

%\title [mode = title]{Enhancing M\&A Predictive Modeling from patent data: A Sapling Similarity adaptation}               

\title [mode = title]{Machine learning-based similarity measure to forecast M\&A from patent data} 

\tnotemark[1,2]

%\tnotetext[1]{This document is the results of the research
%   project funded by the National Science Foundation.}

%\tnotetext[2]{The second title footnote which is a longer text matter
%   to fill through the whole text width and overflow into
%   another line in the footnotes area of the first page.}

\author[1]{G Albora}[orcid=0000-0001-8154-1607]


%\ead{cvr_1@tug.org.in}

%\ead[url]{www.cvr.cc, cvr@sayahna.org}

%\credit{Conceptualization of this study, Methodology, Software}

\author[1,2]{M Straccamore}[orcid=0000-0002-3351-2323]

\cormark[1]

\author[3,1]{A Zaccaria}[orcid=0000-0002-4478-3292]

%\ead{cvr3@sayahna.org}
%\ead[URL]{www.sayahna.org}

%\credit{Data curation, Writing - Original draft preparation}

\affiliation[1]{organization={Centro Ricerche Enrico Fermi},
    addressline={Via Panisperna 89/A}, 
    city={Rome},
    % citysep={}, % Uncomment if no comma needed between city and postcode
    postcode={00184 RM}, 
    country={Italy}}

\affiliation[2]{organization={Sony CSL - Rome, Joint Initiative CREF-SONY},
    addressline={Centro Ricerche Enrico Fermi, Via Panisperna 89/A}, 
    city={Rome},
    % citysep={}, % Uncomment if no comma needed between city and postcode
    postcode={00184 RM}, 
    country={Italy}}

\affiliation[3]{organization={Istituto dei Sistemi Complessi (ISC) - CNR UoS Sapienza},
    addressline={P.le A. Moro, 2}, 
    city={Rome},
    % citysep={}, % Uncomment if no comma needed between city and postcode
    postcode={00185 RM}, 
    country={Italy}}

\cortext[cor1]{Corresponding author}

\begin{abstract}
Defining and finalizing Mergers and Acquisitions (M$\&$A) requires complex human skills, which makes it very hard to automatically find the best partner or predict which firms will make a deal. In this work, we propose the MASS algorithm, a specifically designed measure of similarity between companies and we apply it to patenting activity data to forecast M$\&$A deals. MASS is based on an extreme simplification of tree-based machine learning algorithms and naturally incorporates intuitive criteria for deals; as such, it is fully interpretable and explainable. By applying MASS to the Zephyr
and Crunchbase datasets, we show that it outperforms LightGCN, a "black box" graph convolutional network algorithm. When similar companies have disjoint patenting activities, on the contrary, LightGCN turns out to be the most effective algorithm.
This study provides a simple and powerful tool to model and predict M$\&$A deals, offering valuable insights to managers and practitioners for informed decision-making.

\end{abstract}

% \includegraphics{figs/grabs.pdf}

%\item Research highlights item 1
%\item Research highlights item 2
%\item Research highlights item 3

% Keywords
% Each keyword is separated by \sep

\maketitle

\section{Introduction}
In today's rapidly evolving landscape of technological advancements, companies face the constant challenge of staying at the forefront of innovation. While internal research and development efforts play a significant role, they may not always be sufficient in terms of time and costs to keep up with the swiftly changing technological environment. As a result, many firms seek to expand their technological horizons by engaging in Mergers and Acquisitions (M\&A) \cite{bruner2004applied}. Such operations are used extensively as a financial instrument by firms of any region and size and constitute a business that, only in 2019, has almost reached 4 trillion dollars (\url{https://imaa-institute.org}). Such strategic moves allow companies to tap into the innovation capabilities of their target entities, leverage their technologies, and potentially venture into new markets.
The choice of the possible best target for a deal is made in a complex evolving landscape of partners and competitors, involving a huge effort in terms of time and human capabilities. In this paper, we propose an automatized, machine learning-inspired approach to quantify the closeness between two firms in terms of their patenting activities, and we test this and other measures in an out-of-sample forecast exercise. Equipped with this tool, decision-makers can assess to what extent to exploit a technology sector a firm already masters or explore new innovation possibilities. In order to build a quantitative measure of the similarity between companies, we draw inspiration from the Economic Complexity framework \cite{hidalgo2009building}. In particular, our investigation centers on the concept of "Relatedness" (\cite{hidalgo2018principle, hidalgo2007product, zaccaria2014taxonomy}), which in our study serves as a measure of the similarity between two firms based on the technological sectors found in their patents. Our similarity metric allows us to compare and contrast the patent portfolios of acquiring and target companies, enabling a deeper understanding of the technological dynamics at play in these strategic transactions.

Similarity metrics, such as cosine similarity, are the key to constructing collaborative filtering \cite{schafer2007collaborative}, which is a widely employed technique in recommender systems and link prediction exercises. Recently, in the field of unweighted bipartite networks, it has been introduced a novel metric known as Sapling Similarity \cite{albora2023sapling}. This metric has demonstrated superior performance in link prediction and recommendation tasks compared to existing metrics in the literature. In this study, we have modified the Sapling Similarity to predict M\&A events, introducing the MASS approach. First, MASS correctly considers weighted bipartite networks, which is the context of our firm-technology network. Second,
similarity measures usually assume symmetry, meaning that the probability of firm $f_1$ acquiring firm $f_2$ is the same as that of firm $f_2$ acquiring firm $f_1$. However, this assumption does not hold in the real world, and MASS considers that it is more likely that a large, established firm acquires a small startup than the other way around. Finally, we included a preference of acquirer firms for rare technologies. The simplicity of the mathematical expression of MASS makes its output (the likelihood of M\&A events between two firms) fully interpretable and explainable. Our findings reveal that this approach yields a significant enhancement in our ability to understand and make predictions on future M\&As with respect to other methods, including black-box machine learning such as decision tree-based algorithms \cite{breiman2001random,chen2016xgboost} and graph convolutional networks \cite{he2020lightgcn}.\\
Furthermore, we have delved deeper into our study by considering the distinctions among firms belonging to different sectors. Certain sectors, such as the pharmaceutical industry, tend to generate patents more frequently, whereas others, like the financial sector, exhibit less frequent patenting activity \cite{arundel1998percentage, chabchoub2005explaining,arsini2023prediction}. Notably, when firms possess a low degree in the firm-technology bipartite network, it is common to observe M\&A transactions between two firms that have zero co-occurrences, indicating no shared technological codes. In such instances, all traditional similarity metrics, like Cosine Similarity, but also Sapling Similarity fail to capture any meaningful signals. Our investigation reveals that machine learning-based approaches can deal with events between firms with zero co-occurrences, as they can discern signals that simpler similarity metrics are incapable of detecting.\\
The paper is organized as follows. In section 2 we review the related literature both in M\&A studies and in the Economic Complexity field. In section 3 we briefly discuss our objectives. Section 4 is devoted to the description of our database and methodologies. We present our results about deals prediction in section 5. Section 6 concludes.

\section{Literature Review}
\subsection{Literature Review on M\&A}
Recently, the M\&A literature has witnessed significant expansion, delving into diverse directions. Liu et al. \cite{liu2022influencing} offer guidance on M\&A risks through success and failure cases. Another study \cite{satapathy2022effect} assesses efficiency indicators for acquiring organizations, identifying determinants of acquiring firms based on a sample of ten listed companies that underwent M\&A between 2010 and 2017. Investigating cross-border M\&As, \cite{ding2022effects} explores the differential effects of foreign investor protection (FIP) and domestic investor protection (DIP), finding stronger effects for larger M\&As, target countries further from home countries, and those with better financial development. Additionally, \cite{kooli2021impact} examines the impact of the COVID-19 pandemic on global M\&A activity.

In emerging markets, Ogendo and Ariemba \cite{ogendo2022mergers} propose that firms can utilize mergers and acquisitions during economic downturns to deliver superior value to shareholders. In the context of the Indonesia Stock Exchange, Novita and Rasyid \cite{novita2022comparative} analyze the financial performance of companies before and after M\&A, revealing differences in Return on Assets and Price Earning Ratios but no significant differences in other financial ratios.

A prominent concept in M\&A research is "absorptive capacity" \cite{cohen1990absorptive}, denoting the acquirer company's ability to assimilate knowledge and competencies from the target firm. Building on this, the relatedness between the acquirer and target companies is deemed crucial for successful integration \cite{lane1998relative}. Geographical distance has been found to negatively influence the probability of M\&A occurrences \cite{kaul2016capabilities, chakrabarti2016role}, while similarities in ownership or industrial sector can increase such probabilities \cite{bettinazzi2020ownership, kennedy2002matching}. Analyzing a large set of acquisitions using a similarity measure introduced by Teece et al. \cite{teece1994understanding}, \cite{cefis2013theimportance} establishes a statistically significant correlation between M\&A occurrence and industry relatedness.

A notable section of M\&A studies focuses on the patenting activities of involved firms, centring on technological relatedness. Ahuja-Katilia \cite{AhujaKatilia2001} introduces a measure of technological similarity between acquirer and target firms, revealing an inverse parabolic relationship between technological similarity and innovation performance post-acquisition. Many subsequent authors develop different measures of technological relatedness and investigate their links to post-acquisition performance \cite{Cloodt2006, CASSIMAN2005195, Hagedoorn2002, valentini2010aa, Jo:2016aa, makri2010, orsi2015}. However, results regarding the inverse parabolic behaviour between relatedness and performance are inconclusive due to the lack of a standardized and recognized method for robust performance and relatedness measurements \cite{Jo:2016aa, cimini2022meta}.

While the majority of M\&A literature focuses on correlating relatedness measures with successive performances, \cite{albora2023product} highlights the importance of forecasting as a crucial test for the validity of relatedness assessments. Notable forecast exercises involve \cite{wei2008patent}, which uses an ensemble learning algorithm to predict future acquisitions based on relative features between companies and patent data, and \cite{futagami2021pairwise}, where a tree-based machine learning algorithm is trained on a large set of M\&A features encompassing financial, geographical, industrial, and patent data of firms for M\&A prediction. Finally, in \cite{arsini2023prediction}, the authors develop a method to predict future acquisitions by assuming that companies deal more frequently with technologically related ones.

\subsection{Literature Review on Economic Complexity}
%Economic Complexity is ...
%The starting point is the bipartite network country-product, but there are studies also with technologies, firms, regions, cities, ...
%Relatedness is a fundamental tool and there exists several studies.
%Methods based on cooccurrences, and also machine learning.
%Since the affinity with recommender system LightGCN.
%We adapt these methods to works with the case in which we make recommendations on the same layer.

Economic Complexity (EC) is a conceptual framework that studies the knowledge intensity of an economy or an industrial sector by using network-based and machine-learning approaches. One line of research
analyzes the diversity and sophistication of the productive outputs of countries \cite{tacchella2012new, hausmann2014atlas}, quantifying the varying degrees of economic development among countries, and providing insights into the capabilities of economies and their growth potential. A second line of research stems from the concept of relatedness \cite{hidalgo2018principle}, which measures the similarity between economic activities or the affinity between an economic actor or a geographical area and such activities, for instance exporting a product or patenting in a specific technological field. Measures such as Product Space \cite{hidalgo2007product} and the Taxonomy Network \cite{zaccaria2014taxonomy} have become essential tools for policymakers and economists to understand and predict economic dynamics \cite{diodato2023economic}.
Innovation, usually measured from patent data, serves as a critical driver of economic complexity. The correlation between a country's patenting activity and its industrial and scientific development has been extensively documented \cite{pugliese2019unfolding}. Patents are a proxy for the technological content of an economy, with patent networks often revealing the intricate web of knowledge exchange and influence.
Studies extend beyond the national level to include regions \cite{oneale2021structure, napolitano2018technology, dettmann2013determinants, tavassoli2014role, colombelli2014emergence} and cities \cite{straccamore2023urban, straccamore2023geography, kogler2018patent, kogler2013mapping, balland2015technological}, with findings at the firm level reflecting similar patterns \cite{penrose1959the, teece1994understanding}. Firms' strategies for diversification into new products \cite{albora2022machine} and technologies \cite{straccamore2022will, pugliese2019coherent} often mirror the complexity observed in EC analyses, highlighting the scalability of EC principles from macro to micro levels.
The concept of 'Relatedness' in EC literature has been particularly instrumental in understanding the dynamics of mergers and acquisitions (M\&A). Studies have leveraged this concept to predict the likelihood of M\&A by examining the technological and product similarities between firms \cite{arsini2023prediction, kaneko2022novelty, lane1998relative, kaul2016capabilities, chakrabarti2016role, wei2008patent, futagami2021pairwise}, providing a nuanced view of strategic business decisions.
Recent advances have applied EC principles to predictive modelling, with several studies achieving notable success in forecasting economic indicators \cite{straccamore2022will, albora2022machine, albora2023product, albora2023sapling}. However, the application of EC to M\&A prediction is an innovative frontier that remains underexplored.
The intersection of EC measures and recommender systems has opened new research avenues. Adapting algorithms from recommender systems, researchers have begun to address the unique challenges of M\&A prediction within the EC framework, conceptualizing firms as both the users and items of a recommendation engine.
Despite these advancements, significant gaps remain in the literature, particularly concerning the application of EC to M\&A predictive modelling using patent data. This review highlights the need for a more granular approach capable of considering the nuances of firm-level data and the specificities of technological sectors. This is the main objective of this paper.\\
We conclude this section by stressing the need for scientific validation to confirm or falsify the proposed measures. In the economic complexity field, many different or similar relatedness measures are available, and even small variations can produce very different outputs \cite{cimini2022meta}. As a consequence, an out-of-sample forecasting scheme has been introduced \cite{albora2023product} to compare relatedness measures by using the assumption that economic actors will, on average, engage more frequently in more affine, or related, activities. The results point out the need to use tree-based machine learning methods to quantify relatedness \cite{tacchella2023relatedness}. Sapling Similarity \cite{albora2023sapling} extremely simplifies these approaches, making them fully interpretable and explainable, and keeping the accuracy of machine learning forecasts. In this paper, we will follow the same approach, by suitably modifying Sapling Similarity for M\&A predictions and by verifying the goodness of our approach by checking the out-of-sample forecasting performance of this in comparison with other possible measures.

%This literature review has traversed the evolution of Economic Complexity and its application from a macroeconomic lens down to the level of individual firms. By integrating these insights, this study contributes to the emerging body of research applying EC to M\&A prediction, leveraging patent data to reveal underlying patterns in technological capabilities and strategic business manoeuvres.

\section{Objectives and contribution}
In this study, we address the problem of forecasting mergers and acquisitions deals by analyzing the patenting activities of firms by using network-based and machine-learning approaches borrowed from the Economic Complexity and the recommender system literature. Our purpose is to shed light on the underlying patterns and strategic motivations behind such transactions also contributing to the ongoing discussion on the intricate relationships between technological capabilities, diversification strategies, and corporate performance.

The main contributions of this paper are as follows:

\begin{enumerate}
\item Through the exploration of the relationship between patent data and M\&A, we advance our understanding of the dynamics that shape the innovative landscape of modern businesses.
\item We introduce the MASS algorithm, which represents a generalization of the Sapling Similarity that can take continuous values as input, and takes into account the different sizes of the acquirer and target company and the likely seek of rare technologies.
\item We use the MASS algorithm to forecast new deals, finding that it outperforms other approaches in both predicting new acquirer-target couples and future targets given an acquirer.
\item In specific cases where MASS is hardly applicable, graph convolutional neural networks represent the best option to assess the likelihood of a deal.
\item This study represents an advancement in the general field of recommender systems, as the majority of recommendations typically involve connecting nodes from different layers of the bipartite input network. In other words, usually, in recommender systems, the input is a bipartite user-item network, and items are recommended to users. Here, by comparing the respective portfolios of items (i.e., the technology codes relative to the patenting activities), we recommend users to users (i.e., we predict which firm will make a deal with which firm).  
\end{enumerate}

\section{Data and Methodology}

\subsection{Data}
The research was conducted using and matching four databases. From the PATSTAT database \url{www.epo.org/searching-for-patents/business/patstat} it is possible to obtain information about patent and technology codes; AMADEUS \url{https://login.bvdinfo.com/R0/AmadeusNeo} is used to associate patent to companies; finally, Zephyr \url{https://www.bvdinfo.com/it-it/le-nostre-soluzioni/dati/greenfield-investment-and-ma/zephyr} and Crunchbase database (\url{www.crunchbase.com}) are used to find information about M\&A deals.
%The companies’ industrial sectors are obtained from the Crunchbase database.
In this section, we will further describe the preprocessing of our dataset and the method to construct the database we used to build our similarity measures and to perform the forecast exercise.

\subsubsection{Patent data}
The data on technologies and patents utilized in this research is sourced from the Worldwide Patent Statistical Database (PATSTAT, \url{https://www.epo.org/searching-for-patents/business/patstat.html}), maintained by the European Patent Office (EPO). This comprehensive database compiles and organizes information from various regional and national patent offices. A crucial feature of PATSTAT for our study is its incorporation of the International Patent Classification (IPC) system. The IPC, an internationally recognized hierarchical classification system, is regularly updated and managed by the World International Patent Organization (WIPO). It employs a standardized coding system to categorize patents based on technological aspects. These codes are structured into levels of increasing generality, starting from over 70,000 specific groups at the most detailed level to just 8 broad sections at the highest level.
For instance, the code prefix "A" denotes the broad category of "Human Needs," while "C" indicates "Chemistry." Delving deeper, the code "A01" specifies the "Agriculture; Hunting" sector, and "A43" pertains to the "Footwear" sector. It is pertinent to mention that this study excludes the "99" classes and "Z" subclasses. These categories are reserved for technologies that do not fit into the predefined classes or subclasses, rendering them undefined for our analysis.
Further details on this dataset and its application can be found in the study by \cite{pugliese2019coherent}, which provides an in-depth exploration of the subject.

\subsubsection{Firm data}
Data on companies was sourced from the AMADEUS database (\url{https://login.bvdinfo.com/R0/amadeusneo}), which archives information on over 20 million companies, predominantly located in Europe. The AMADEUS database is curated by Bureau van Dijk Electronic Publishing (BvD), a firm renowned for its comprehensive financial, administrative, and budgetary data on corporations. A significant feature of BvD's management is the incorporation of patent identifiers that align with those used by the European Patent Office. This alignment ensures compatibility between the AMADEUS and PATSTAT databases, facilitating integrated analyses \cite{pugliese2019coherent}.
It is acknowledged that AMADEUS has a notable limitation in its coverage, with large corporations being comprehensively represented, whereas smaller entities, particularly those with fewer than 20 employees, are less thoroughly documented \cite{ribeiro2010oecd}. However, for the objectives of this study, this discrepancy does not pose a significant issue.

\subsubsection{Dataset from Crunchbase and Zephyr for M\&A}
Data on M\&A were sourced from two distinct databases: Zephyr and Crunchbase. Zephyr (\url{https://www.bvdinfo.com/en-us/our-products/data/greenfield-investment-and-ma/zephyr}), a commercial database managed by Bureau van Dijk Electronic Publishing (BvD), provides detailed records on M\&A activities, Initial Public Offerings (IPOs), Private Equity, Venture Capital initiatives, and associated rumours on a global scale. For this study, we specifically utilized Zephyr's segment dedicated to entities within the biopharmaceutical industry. This segment encompasses data on approximately 4,000 transactions spanning from 1997 to 2016, involving over 3,700 companies.
Crunchbase (\url{https://www.crunchbase.com}), another commercial repository, was initially established to monitor start-ups. It offers extensive data on both public and private companies, including details on acquisitions, mergers, and investments worldwide. Compared to Zephyr, Crunchbase's database is significantly larger, containing records of more than 100,000 acquisitions dating back to 1922, and information on over a million companies.

\subsubsection{Dataset creation}
The construction of our dataset began with the AMADEUS-Patstat database, which creates a bipartite network of companies linked to technology codes from their patents, as elucidated in \cite{pugliese2019coherent}. Companies are identified by BVDID, which correlates with their patent technology codes by weight, indicating patent share per technology.
Incorporating the Zephyr dataset, we directly mapped 430 companies to their technological profiles from an initial pool of 3167 M\&A-involved entities. For the Crunchbase dataset, a name-cleaning algorithm was employed to match company names to their BVDIDs, culminating in 12017 companies being linked to appropriate technological portfolios out of 28137 candidates. Where multiple BVDIDs emerged, typically for multinational entities with various subsidiaries, we consolidated the corresponding technological portfolios. The M\&A analysis, confined to the period between 2002 and 2012 and to companies with patenting activity from 2000, yielded a sample of 1279 M\&A events across 1974 companies.
Crunchbase's proprietary industrial sector taxonomy, featuring 744 categories and 43 category groups, was refined into 13 aggregated sectors. This reclassification allowed us to distinctly categorize 8069 firms, which represents approximately 70\% of the companies aligned with their technological portfolios.
For our analyses, we selected a subset of the M\&A dataset that exclusively includes companies with a singular sector designation based on our refined classification. This subset comprises 8737 companies, with 913 participating in 547 M\&A transactions.
The temporal aspect of the dataset is encapsulated in 13 yearly adjacency matrices, $\textbf{M}^y$, spanning 2000 to 2012. These matrices chart the relationships between 8,737 companies and 7,132 technologies, with each element $M_{ft}^y$ signifying the affiliation of a firm $f$ with a technology $t$ for a specific year $y$. The matrices are constructed by assigning a uniform weight to each patent, distributed among all pertinent firm-technology pairs and aggregated annually. This method recognizes that patents may cover several technologies and are rarely filed by multiple firms.
We extend the analysis by considering cumulative matrices $\textbf{M}^Y$, each summarizing the technological involvement of firms from 2000 to year $Y$, to reflect a firm's evolving innovation profile. These cumulative matrices underpin our predictive models, which hypothesize that the similarity in technological portfolios between companies can forecast potential M\&A activities in year $Y$.

The resultant dataset, employed in the subsequent analyses of this paper, features 8,737 companies, of which 913 were involved in 547 M\&A deals. The companies were selected based on the availability of a unique industrial sector from the Crunchbase data, enhancing the precision of our predictive modeling.

\subsection{Methods}
In this section, we describe the various similarity metrics used to assess the similarity of companies from their patents. The starting point is the matrix $\textbf{M}^Y$ whose element $M_{ft}$ is the number of patents firms $f$ submitted in the technological sector $t$ during year $Y$. In the following, we will omit the year specification to lighten the notation. This matrix is the representation of a bipartite firm-technology network, and the matrix elements quantify the weights of such a network. Given an acquirer firm A and a target firm T, we want to compute $B_{AT}$, the likelihood of firm A acquiring firm T; different methods will provide different estimations of $B_{AT}$. We will then use these assessments to predict M\&A deals.

\subsubsection{Notation}
In the following, we will define the matrix $\Lambda$ as the result of the scalar product between $\textbf{M}$ and its transpose, $M\cdot M^T$. This matrix is square, and each row represents a possible acquirer company, while each column represents a possible target company. The matrix element $\Lambda_{AT}$ represents the scalar product between the row vectors A and T of the $\textbf{M}$ matrix. Note that if $\textbf{M}$ is binary, $\Lambda_{AT}$ is equal to the number of co-occurrences $CO_{AT} = \sum_{\lambda}M_{A\lambda}M_{T\lambda}$, i.e. the number of technologies firms $A$ and $T$ share. \\ We will denote the row vector of $\Lambda$ corresponding to firm A as $\Lambda_{(A)}$ and the column vector of $\Lambda$ corresponding to firm T as $\Lambda^{(T)}$. Finally, we will utilize the notation $max(\Lambda_{(A)})$ to denote the maximum element of the $\Lambda_{(A)}$ vector, $max(\Lambda^{(T)})$ to denote the maximum element of the $\Lambda^{(T)}$ vector and $max(\Lambda)$ to denote the maximum element of the matrix $\Lambda$.\\

\subsubsection{Jaffe similarity}
The first measure we introduce is Jaffe similarity, introduced in 1986 by Jaffe to quantify the productivity of manufacturing R\&D \cite{jaffe1986technological}. Since then, in the work of Valentini and Dawson \cite{valentini2010beyond}, the Jaffe measure was applied in the M\&A context.
More recently, \cite{arsini2023prediction} conducted a comparative analysis of various methodologies, including machine learning algorithms, for estimating the likelihood of a deal. They concluded that the most effective approach involves assessing the similarity in technological portfolios between the two companies, utilizing Jaffe or cosine similarity as the metric for similarity. In our notation, the equation for Jaffe similarity reads
\begin{equation}
B^{Jaf}_{AT} = \frac{\sum_\lambda{M_{A\lambda}M_{T\lambda}}}{\sqrt{\sum_\lambda{M_{A\lambda}^2}}\sqrt{\sum_\lambda{M_{T\lambda}^2}}}
= \frac{\Lambda_{AT}}{\sqrt{\sum_\lambda{M_{A\lambda}^2}}\sqrt{\sum_\lambda{M_{T\lambda}^2}}}
%\xrightarrow{\text{unweighted}}\frac{CO_{AT}}{\sqrt{k_Ak_T}}
\end{equation}

\subsubsection{Sapling Similarity for unweighted bipartite networks}

\begin{figure*}[h!]
\centering
\includegraphics[width=\textwidth]{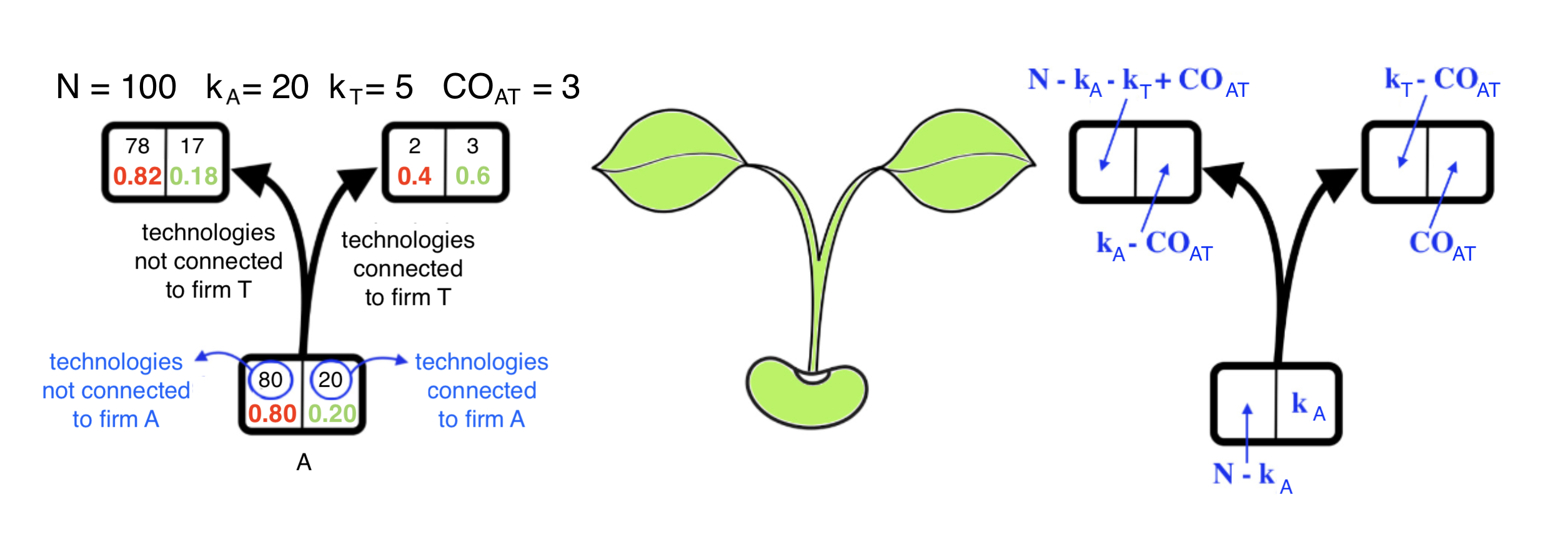}
\caption{On the left figure, an example of a decision sapling illustrating the relationship between two firms, $A$ and $T$, in a scenario featuring $100$ technologies. The node at the bottom indicates the percentage of technologies that are (right) and are not (left) connected to firm A; in this figure, A is connected to $20\%$ of technologies, as denoted by the green value. The upper nodes show how these percentages change when considering only technologies connected to Firm T (right node) or only those not connected to Firm T (left node). On the right figure, we show the value of each box as a function of the number of co-occurrences CO, the degrees k, and the total number of technologies $N$.}
\label{fig:sapling}
\end{figure*}

\cite{albora2023sapling} has recently introduced a new metric of similarity between nodes in unweighted bipartite networks: the Sapling Similarity. The idea behind this metric is to extract the main ingredients of tree-based machine learning models (since they outperform other approaches in Relatedness estimations, see \cite{tacchella2023relatedness}), to allow full interpretability and explainability while preserving the prediction performance. Sapling Similarity is based on the computation of the variation of the Gini impurity (GI) in a decision tree with depth 1, also called decision sapling.
Figure \ref{fig:sapling} shows a decision sapling. 
On the left, we represent a numerical example in which only $N=100$ technologies exist and two firms, A and T, are considered. A is connected to (i.e., has patenting activity in) $k_A$=20 technologies, and T is connected to $k_T$=5 technologies. A and T share $CO_{AT}$=3 technologies. The bottom node indicates the fraction of the technologies that firm A is either connected with or not; notably, firm $A$ has established connections with $20\%$ of the technologies, as denoted by the green value. Upper nodes detail the updates of these proportions when the analysis is narrowed to technologies either connected with or not connected with firm $T$. In this example, knowing that a technology is connected with T increases the probability that it is connected to A from 20\% to 60\%, suggesting a positive similarity between A and T.

To quantify this line of reasoning, we need a measure of the polarization in a node.
Given $p_1$ the fraction of positive samples in a decision tree node, and $p_0$ the fraction of negative samples, the Gini Impurity of the node is defined as:
\begin{equation}
GI = 1-p_0^2-p_1^2 = 2p_0p_1.
\end{equation}
This quantity tells us how much the samples in the node are peaked, or concentrated, towards the positive or the negative case: maximum polarization (only positive or negative samples) implies $GI=0$ while minimum polarization (equal number of positive and negative samples) means $GI = 0.5$.
So if we consider the lower node of the decision sapling, we have $p_0 = 0.8$, $p_1 = 0.2$, and $GI^{low} = 0.32$. Analogously we find that the Gini Impurity of the upright node is $GI^{upr} = 0.48$ and the one of the upleft node is $GI^{upl} = 0.29$. The variation of the Gini impurity is defined as:
\begin{equation}
\frac{\Delta GI}{GI^{low}} = \frac{GI^{low}-f^{upl}GI^{upl}-f^{upr}GI^{upr}}{GI^{low}}
\end{equation}
where $f^{upl}$ and $f^{upr}$ are the fraction of samples that are respectively in the upper left and the upper right node (in the case of the figure, 0.95 and 0.05).\\
The variation of the Gini Impurity quantifies how much the information that T patents or not patents in a generic technology field $\lambda$ is important to understand whether A patents in $\lambda$. This is the absolute value of the Sapling Similarity; its sign is positive if $p_1^{upr}\geq p_1^{low}$ (which means that knowing that T is connected to $\lambda$ increases the probability that also A is connected to it), and negative otherwise.\\
Using the figure on the right, in which we report the general formulas, we can easily derive an equation for the sapling similarity as a function of the co-occurrences $CO_{AT}$ (how many technologies firms A and T share), the degrees $k_A$ and $k_T$ (the number of technologies each firm has) and the total number of technologies in the bipartite network $N$:

\begin{equation}
	B_{AT}^{sap} =
	\begin{cases}
		1-f_{AT} ~\text{ if } \frac{CO_{AT}  N}{k_{A}k_{T}}\geq1\\[3mm]
		-1+f_{AT} ~\text{ otherwise }
	\end{cases}	
\label{eq:sapling1}
\end{equation}
Where:
\begin{equation}
f_{AT} = \frac{CO_{AT}\left(1-\frac{CO_{AT}}{k_{T}}\right)+\left(k_{A}-CO_{AT}\right)\left(1-\frac{k_{A}-CO_{AT}}{N-k_{T}}\right)}
{k_{A}\left(1-\frac{k_{A}}{N}\right)}.
\label{eq:sapling2}
\end{equation}

\subsubsection{Generalization of the Sapling Similarity for weighted bipartite networks}
In our specific case, the input data is a bipartite network that connects firms to the technology fields, which is weighted; so we need to generalize the Sapling Similarity, which in previous papers has always been used with binary inputs. Our line of reasoning goes as follows. The degree $k_F$ of a firm $F$ can be seen as the maximum number of co-occurrences that this firm can have (indeed, it cannot share with another firm a number of technologies larger than the number of its technologies); on the other hand, we can think of N as the maximum number of co-occurrences that two generic firms can have (that is, the number of co-occurrences between two firms that possess all the technologies).\\
In the case in which the matrix $\textbf{M}$ takes continuous values, the number of co-occurrences between the two firms A and T is generalized as the scalar product between the two row-vectors of the matrix $\textbf{M}$ that represent the patenting activity of A and T.
\begin{equation}
CO_{AT}\xrightarrow{\text{continuous}}M_A\cdot M_T = \sum_\lambda{M_{A\lambda}M_{T\lambda}} \overset{\text{def}}{=} \Lambda_{AT}
\label{eq:cont}
\end{equation}
To simplify the following equations we introduce the matrix $\Lambda$ whose elements are defined in equation \ref{eq:cont}.
Since the elements of $\textbf{M}$ take continuous values without an upper limit, in principle the theoretical maximum value of $\Lambda_{AT}$ is infinite. However, we can take this value from the empirical counterpart: so we consider $\max(\Lambda_{(A)})$ as the equivalent of $k_A$ and $\max(\Lambda^{(T)})$ as the equivalent of $k_T$. With the same reasoning, we can say that the continuous equivalent of $N$ is $\max(\Lambda)$. So with these changes, the equation of the sapling similarity in the continuous case reads:

\begin{equation}
	B_{AT}^{sap} =
	\begin{cases}
		1-f_{AT} ~\text{ if } \frac{\Lambda_{AT}\max(\Lambda)}{\max(\Lambda_{(A)})\max(\Lambda^{(T)})}\geq1\\[3mm]
		-1+f_{AT} ~\text{ otherwise }
	\end{cases}	
\label{eq:saplingC1}
\end{equation}
Where:
\begin{equation}
f_{AT} = \frac{\Lambda_{AT}\left(1-\frac{\Lambda_{AT}}{max\left(\Lambda^{(T)}\right)}\right)+\left(\max\left(\Lambda_{(A)}\right)-\Lambda_{AT}\right)\left(1-\frac{\max\left(\Lambda_{(A)}\right)-\Lambda_{AT}}{\max\left(\Lambda\right)-\max\left(\Lambda^{(T)}\right)}\right)}{\max\left(\Lambda_{(A)}\right)\left(1-\frac{\max\left(\Lambda_{(A)}\right)}{\max\left(\Lambda\right)}\right)}
\label{eq:saplingC2}
\end{equation}

\subsubsection{Considering firms' size and technologies' ubiquity}
\begin{figure*}[h!]
\centering
\includegraphics[width=0.9\textwidth]{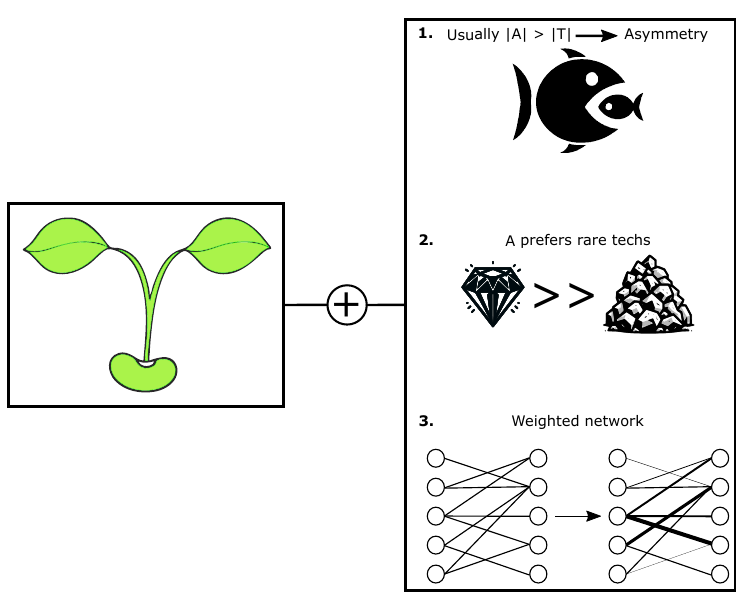}
\caption{Overview of the MASS method. The starting point is the Sapling Similarity measure, schematically depicted on the left. In order to predict M\&A deals, three modifications have been implemented: (1) the consideration of acquirer and target firm sizes, (2) an increased emphasis on co-occurrences involving rare technologies, and (3) the generalization to continuous input variables (that is, weighted bipartite networks).}
\label{fig:res_pred}
\end{figure*}

The generalization of the Sapling Similarity discussed above allows one to apply it to continuous variables. To obtain the MASS similarity measure, which is optimally designed to forecast M\&A deals, we need to include two further features. The three changes are schematically represented in Figure \ref{fig:res_pred}.\\
In the literature (see \cite{kaneko2022novelty, AhujaKatilia2001, cefis2013theimportance, Cloodt2006, arsini2023prediction}) the similarity metrics used to estimate the probability of a firm $A$ acquiring a firm $T$ are symmetric. In other words, the estimated probability that $A$ acquires $T$ is the same that $T$ acquires $A$. However, this does not correspond to reality: the probability that a huge firm acquires a small firm is not the same that a small firm acquires a huge one. To take this effect into account we multiply the value of the continuous Sapling Similarity between two firms $A$ and $T$ by a factor that takes into account the size of the firms, that we proxy using the norm of their portfolios:
\begin{equation}
	B_{AT}^{sap (1)} =\frac{\sqrt{\sum_\lambda{M_{A\lambda}^2}}}{\sqrt{\sum_\lambda{M_{T\lambda}^2}}}B_{AT}^{sap}
\label{eq:mod1}
\end{equation}
In this way, if the size of the acquirer is larger than the size of the target the probability that A acquires $T$ will be larger than the one that $T$ acquires $A$.\\
The second effect we want to include is based on the idea that if two firms share a rare technology (a technology sector in which only a few firms patent), then it is more probable to have an acquisition between these two firms. Let us identify a simple technology with rocks and a complex technology with diamonds, as in Figure \ref{fig:res_pred}. A firm that produces both rocks and diamonds can increase the production of rocks without particular effort because rocks are very simple and common. However, if the same firm wants to increase the production of diamonds, a complex or rare product, that firm could consider acquiring another firm that already specializes in diamonds. To consider this effect in our similarity measure we define $v_\lambda = \sqrt{\sum_f{M_{f\lambda}^2}}$ and when we compute the scalar product between the two-row vectors of $\textbf{M}$, we add a factor $v_\lambda$ in the denominator, defining in this way $\tilde{\Lambda}$:
\begin{equation}
    \Lambda_{AT} = M_A\cdot M_T \rightarrow \tilde{\Lambda}_{AT}\overset{\text{def}}{=} \sum_\lambda{\frac{M_{A\lambda}M_{T\lambda}}{v_\lambda}}.
    \label{eq:mod2}
\end{equation}
So in equation \ref{eq:saplingC1} and \ref{eq:saplingC2} matrix $\Lambda$ is substituted by $\tilde{\Lambda}$.\\
In the results section, we refer to the term SS as Sapling Similarity, which can be either weighted or unweighted and is utilized without the adjustments (1) and (2) reported in Figure \ref{fig:res_pred}. By SS(1), we denote the inclusion of the modification defined in equation \ref{eq:mod1} that accounts for the size of the firm, while SS(2) refers to the inclusion of modification (2), defined in equation \ref{eq:mod2}, which adjusts for the 'rarity' of the technologies. When we mention SS(1+2), we are referring to the incorporation of both modifications. Finally, the metric we designate as MASS (Mergers and Acquisitions Sapling Similarity) corresponds to the weighted version of SS(1+2).

\subsubsection{LightGCN}
In the context of recommendation systems, Graph Convolutional Networks (GCNs) (\cite{hamilton2017inductive, kipf2016semi}) have become increasingly popular due to their ability to capture complex relationships within data. Essentially, GCNs work by learning features from graph structures, such as networks of users and products, by considering the connections and the features of neighboring nodes.\\
Light Graph Convolutional Network (LightGCN), introduced by \cite{he2020lightgcn}, is a streamlined variant of the traditional GCN, specifically designed for recommendation systems. LightGCN simplifies the GCN architecture by removing feature transformation and non-linear activation functions. This simplification aims to reduce computational complexity while maintaining, or even enhancing, the performance in recommendation tasks.\\
LightGCN operates directly on the user-item interaction graph. It effectively learns user and item embeddings by aggregating features from neighboring nodes, capturing both direct and indirect interactions within the graph. This approach allows LightGCN to efficiently and accurately model the preferences and behaviors of users, leading to improved recommendation quality.\\
In this study, we will use LightGCN to predict M\&A and we will compare its performance with Jaffe Similarity (that outperforms various machine learning approaches, see \cite{arsini2023prediction}), Sapling Similarity, and the newly introduced MASS.
\section{Experiments}
\subsection{Pair, Target, and Acquirer prediction}

In this paper, we aim to compare the effectiveness of various methods in predicting future M\&A deals between companies. We estimate the probability of the possible M\&As occurring between companies in a given year Y using only past data; so the prediction scores are recalculated in each year. To assess the quality of the methodologies, we examine their ability to predict M\&As between companies. In practice, this translates into three distinct binary classification exercises:
\begin{enumerate}
\item Pair Prediction: For this exercise, the 547 pairs of companies that undergo M\&A are labeled as positive events, while negative labels are assigned to randomly generated pairs of companies. For each M\&A that occurs in year Y, 200 negative pairs are generated, ensuring that each pair is unique and not among the 547 actual M\&A pairs. The best-performing model is the one that accurately distinguishes the true M\&A pairs from the randomly generated ones.
\item Target Prediction: Here, for each of the 547 actual acquirers, 200 negative targets are generated ensuring that these do not coincide with the real targets and that there are no repetitions. The optimal model is identified as the one that can effectively differentiate the actual target of the single acquirers from those randomly generated.
\item Acquirer Prediction: Similar to the target prediction, this exercise involves generating 200 negative acquirers for each of the 547 true targets. The model's task is to identify the true acquirer of each target.
\end{enumerate}
In all three types of experiments, for every single M\&A deal, 200 negative company pairs are generated, resulting in a class imbalance of 1:200 in the binary classification exercise. To quantify the performance of the models in these three types of experiments, we use standard performance indicators for binary classification \cite{cunningham2008supervised, caruana2006empirical}. For pair prediction, we employ the following three indicators:
\begin{itemize}
\item Best F1 score (best F1) \cite{tacchella2023relatedness, albora2023product, cruz2016tackling}: This score is computed by finding the threshold that maximizes the F1 score \cite{dice1945measures}, that is defined as the harmonic mean of precision and recall;
\item Area under the Precision-Recall Curve (AUC PR) \cite{boyd2013area, keilwagen2014area}; the area under the curve on the precision-recall plane. This area is derived by varing the threshold that determines the score above which predictions are classified as positive.
\item Precision at 500 (prec@500); To compute this metric, we evaluate the top 500 scoring elements. prec@500 measures the proportion of these top 500 elements that are true positives (accurately predicted positively) out of the total 500 elements examined.
\end{itemize}
For target and acquirer prediction, we utilize:
\begin{itemize}
\item Best F1 score (best F1), as for pair prediction;
\item Hit Ratio at 5 (HR 5) \cite{he2017neural}: it measures the proportion of times that the relevant item (the true acquirer in the acquirer prediction exercise and the true target in the target prediction exercise) appears in the list of the top 5 recommendations;
\item Mean Average Precision (mAP): the mean of the average precision \cite{salton1983introduction} across all acquirers (target prediction exercise) or targets (acquirer prediction exercise);
\end{itemize}
Each experiment is repeated 20 times, with negative M\&As being regenerated each time. The final score is an arithmetic mean of these 20 repetitions.
\subsection{Results}
\begin{figure*}[h!]
\centering
\includegraphics[width=0.9\textwidth]{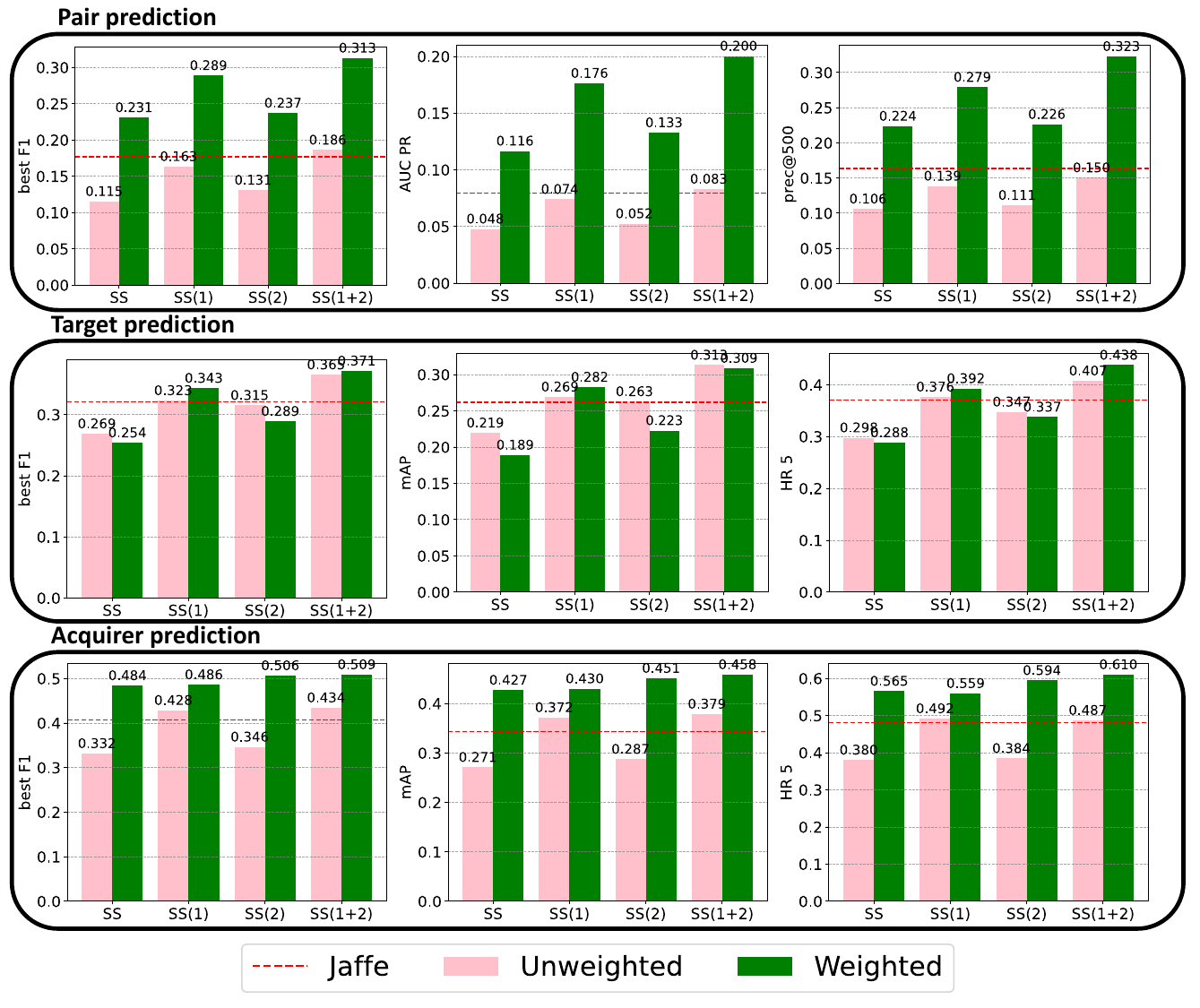}
\caption{Performance of the various Sapling Similarity variants in predicting M\&A deals in the three cases: pair, target, and acquirer prediction. The variants, reported on the x-axis, are described in the Methods section. We report in pink the case in which the input network binary, and in green, the weighted case. The MASS algorithm corresponds to the green SS(1+2) variant. All improvements enhance the prediction performances, MASS being the outperforming algorithm.}
\label{fig:results_sapling}
\end{figure*}

The prediction performances of the different methods are compared in Figure \ref{fig:results_sapling}, which shows nine bar plots arranged in a 3x3 grid. Each row corresponds to a different prediction exercise: pair prediction, target prediction, and acquirer prediction, respectively. Within each row, three distinct performance indicators are utilized for evaluation: In order Best F1 Score, Area Under Precision Recall Curve, and Precision at 500 for pair prediction; Best F1 Score, Mean Average Precision, and Hit Ratio at 5 for both target and acquirer predictions. The bar plots feature eight bars, grouped in pairs, with pink bars representing predictions made using an unweighted network and green bars for those made using a weighted network. Each pair of bars corresponds to a variant of sapling similarity, incorporating modifications (1) and (2), introduced in the methods section, aimed at refining the prediction accuracy for this particular case study. The current state of the art is Jaffe similarity (red dashed line).\\
%\begin{figure*}[h!]
%\centering
%\includegraphics[width=0.3\textwidth]{figs/pair_best_F1.eps}
%\includegraphics[width=0.3\textwidth]{figs/pair_AUC_PR.eps}
%\includegraphics[width=0.3\textwidth]{figs/pair_prec@500.eps}\\
%\includegraphics[width=0.3\textwidth]{figs/target_best_F1.eps}
%\includegraphics[width=0.3\textwidth]{figs/target_mAP.eps}
%\includegraphics[width=0.3\textwidth]{figs/target_HR_5.eps}\\
%\includegraphics[width=0.3\textwidth]{figs/acquirer_best_F1.eps}
%\includegraphics[width=0.3\textwidth]{figs/acquirer_mAP.eps}
%\includegraphics[width=0.3\textwidth]{figs/acquirer_HR_5.eps}
%\caption{Performance of sapling similarity in predicting M\&A in the three cases: pair, target, and acquirer prediction. On the x-axis, the variants are described in the Methods section, and in blue, the case where the network is used without weights, while in orange, is the case where the network is used with weights.}
%\label{fig:results_sapling}
%\end{figure*}

The modifications introduced to the base equation of Sapling Similarity consistently yield improvements. This enhancement aligns with the rationale behind their introduction: smaller target companies are generally the focus of acquisitions, addressed by modification (1), and the attractiveness of a small company possessing a rare technology, as considered in modification (2).\\
Generally, predictions utilizing a weighted network outperform those based on an unweighted network, validating the effectiveness of extending Sapling Similarity to incorporate the case of weighted networks. However, in target prediction scenarios, the distinction between weighted and unweighted networks is minimal. This suggests that for large companies acquiring smaller ones, the mere possession of a technology by the target is significant, irrespective of whether it is the target's main technology or not. In contrast, understanding the importance of technologies to the acquiring company is crucial for accurately predicting the acquirer.\\
The MASS approach, incorporating strategic modifications to the already well-performing Sapling Similarity, consistently outperforms the previous state-of-the-art, that is Jaffe's method (the red dashed line in the plots), which previous studies shown to perform better than a number of other approaches \cite{arsini2023prediction}. This superiority across all metrics and prediction exercises underlines the method's robustness and its capacity to set a new benchmark in the field of M\&A prediction.\\
Given that target prediction and acquirer prediction utilize the same performance indicators, a direct comparison reveals that acquirer prediction exercises generally achieve higher performance metrics. This indicates that predicting the acquirer in a M\&A scenario is more straightforward than identifying the target.

\subsection{Acquisitions with low co-occurrences}
\begin{figure*}[h!]
\centering
\includegraphics[width=0.9\textwidth]{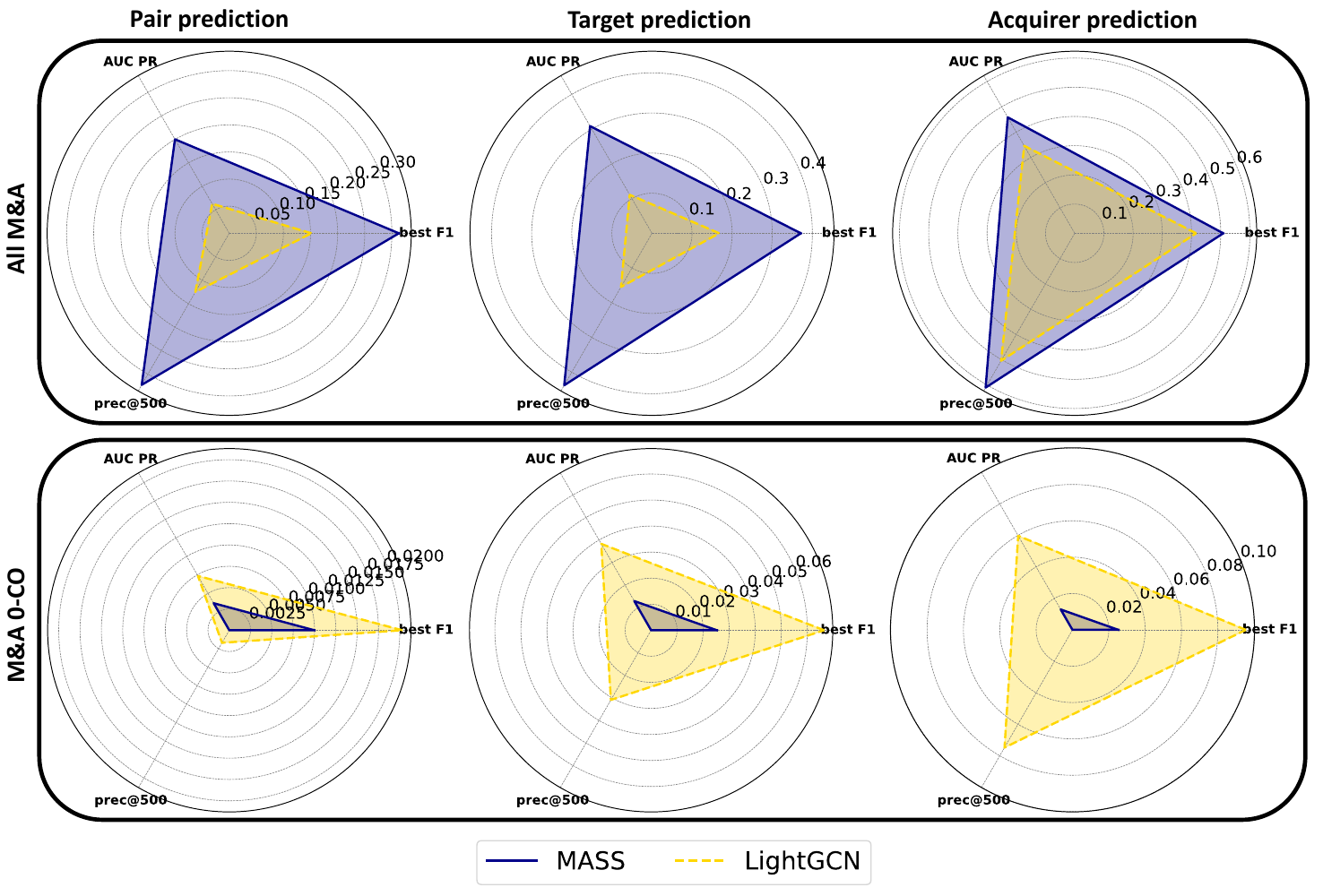}
\caption{Comparative analysis of MASS and LightGCN performances across different M\&A prediction scenarios. The top row of radar plots presents the performance metrics for both methods when the test is conducted on the entire dataset of 547 M\&As. Here, MASS outperforms. The bottom row depicts the same exercises, but the test focuses exclusively on the subset of 123 M\&As between companies with zero co-occurrences (0-CO). In this case, LightGCN is able to capture the hidden, higher-order similarities between companies.}
\label{fig:lightGCN}
\end{figure*}

In the previous section, our investigation has shown the robustness of the MASS approach as the state-of-the-art approach for forecasting M\&As between companies from their patenting activity. However, a pertinent inquiry arises regarding the limitations of this method.\\
Indeed, not all companies exhibit the same propensity for patenting, leading to a significant number of firms with very few technologies - or links in the bipartite network language. This variance can largely be attributed to diverse patenting policies across different sectors. Specifically, certain industries are naturally inclined to generate fewer patents than others. Consequently, some sectors tend to have limited patenting activity, leading to a higher likelihood of M\&A deals occurring between firms with zero co-occurrences. Despite its prowess, methods like MASS struggle to predict these instances due to their reliance on shared technologies for similarity calculation.\\
Among the 547 M\&A instances in our dataset, 123 involve pairs of companies with no direct technology overlap, highlighting a substantial subset where Sapling Similarity for M\&A's predictive power is limited.
This is where machine learning comes to help.\\
In Figure \ref{fig:lightGCN}, we compare the performances of MASS and LightGCN across two different testing scenarios: the entire dataset of 547 M\&As (above). and the subset of 123 M\&As between companies with zero co-occurrences (below). The radar plots schematically represent the prediction ability of the two approaches in the three different exercises, the largest area being relative to a higher performance. We note that in generating random pairs of companies for negative test cases, care was taken to ensure that none of the pairs matched the 547 actual M\&A instances, even when only 123 M\&As are considered. Within these plots, the vertices correspond to the three performance metrics utilized in our evaluation. The resulting shapes formed by connecting these vertices thus visually encapsulate the comparative performance of MASS and LightGCN across these metrics.

%\begin{figure*}[h!]
%\centering
%\includegraphics[width=0.3\textwidth]{figs/radar_pair.png}
%\includegraphics[width=0.3\textwidth]{figs/radar_target.png}
%\includegraphics[width=0.3\textwidth]{figs/radar_acquirer.png}\\
%\includegraphics[width=0.3\textwidth]{figs/radar_pair_0.png}
%\includegraphics[width=0.3\textwidth]{figs/radar_target_0.png}
%\includegraphics[width=0.3\textwidth]{figs/radar_acquirer_0.png}\\
%\caption{Comparative Analysis of Sapling Similarity from M\&As and LightGCN Performances Across Different M\&A Prediction Scenarios. The top row of radar plots presents the performance metrics for both methods when the test is conducted on the entire dataset of 547 M\&As. The bottom row depicts the same but the test focuses exclusively on the subset of 123 M\&As between companies with zero co-occurrences.}
%\label{fig:lightGCN}
%\end{figure*}

The graphical representation clearly illustrates that while MASS outperforms LightGCN across the board when considering the full dataset, its supremacy diminishes in scenarios involving companies with no technology overlap, i.e. with zero co-occurrences (0-CO). In these cases, LightGCN emerges as the superior predictive tool. This shift can be attributed to LightGCN's ability to leverage the broader structure of the bipartite network, capturing latent similarities between companies beyond direct technological co-occurrences. By effectively utilizing this comprehensive network information, LightGCN demonstrates a pronounced advantage in predicting M\&As among companies that, on the surface, share no common technological ground.

\section{Conclusions}

This study provides significant insights into the dynamics of Mergers and Acquisitions (M\&A), utilizing the Economic Complexity framework, and in particular neural networks and machine learning-inspired algorithms. Starting from the bipartite firm-technology network, we estimate the likelihood of an M\&A deal occurring between two firms. A recent study \cite{arsini2023prediction}, using the same data, demonstrated that cosine or Jaffe similarity between firms is a good estimator of this probability, outperforming other similarity measures and even machine learning approaches. In this study, we have outperformed this result by introducing the M\&A Sapling Similarity (MASS) approach. Moving from the Sapling Similarity, a metric recently introduced in \cite{albora2023sapling}, we generalize it to take into account weighted bipartite networks; furthermore, we add two modifications that account respectively for the fact that acquirer firms are usually large and target firms are small and that when counting co-occurrences between two firms, those between rare technologies should weigh more in estimating the probability of a M\&A.\\
The results from three different prediction exercises (pair prediction, target prediction, and acquirer prediction) show that our method represents the state-of-the-art in estimating the probability of a M\&A occurring between two firms.
Furthermore, this study also investigates the case of working with firms that produce few patents. In such cases, it is common to encounter M\&As between firms with zero technologies in common. In these instances, methods like Cosine Similarity and MASS have limited predictive power since they hardly detect a similarity signal between the two firms. In this scenarios, we show that LightGCN, a graph convolutional network introduced in \cite{he2020lightgcn} outperforms other approaches.\\
This paper advances theoretical knowledge on the intersection of Economic Complexity, technological innovation, and M\&A activities; beyond this peculiar application, indeed, the MASS algorithm - or other suitable modifications - represents a first attempt to adapt recommending systems approaches, which are usually relative to bipartite networks - to a monopartite case study. Finally, our work provides actionable insights for practitioners involved in strategic planning and corporate finance.

\section{Data and code availability}
All data and Python code necessary to reproduce the results presented in this paper are available at\\https://github.com/giamba95/SaplingSimilarity/tree/main/m\&a.

\section{Acknowledgements}
We thank Arianna Martinelli and Lorenzo Napolitano for kindly providing data in the context of the CrisisLab-ProCoPe project. We also thank the European Union - Next Generation EU PRIN project no. 20223W2JKJ ”WECARE”.

\printcredits

\bibliographystyle{model1-num-names}
%\bibliographystyle{cas-model2-names}

\bibliography{zzzzz}


\end{document}